\documentclass[aps,prd,superscriptaddress,A4paper,preprintnumbers,nofootinbib]{revtex4}
\usepackage{amssymb,amsmath,graphicx,epsfig,bm,color}
\usepackage{bm}
 \usepackage{float}
\usepackage{slashed}
\usepackage{verbatim} 
\newcommand{\be}{\begin{eqnarray}}
\newcommand{\ee}{\end{eqnarray}}
\usepackage[colorlinks]{hyperref}
\usepackage{leftidx}
\usepackage{booktabs}
\usepackage{latexsym}
\usepackage{revsymb}
\usepackage{multirow}
\usepackage{caption}
\usepackage{subcaption}
\usepackage{comment}
\usepackage[normalem]{ulem}
\hypersetup{
    colorlinks=red,
    citecolor=blue,
    filecolor=magenta,      
    urlcolor=red,
}

\renewcommand{\d}{\mathrm{d}}

\newcommand{\xB }{x_{\scriptscriptstyle B}}
\newcommand{\sT}{{\scriptscriptstyle T}}

\definecolor{dipankar_color}{rgb}{0.0, 0.9, 0.0}
\definecolor{raj_color}{rgb}{0.0, 0.0, 1.0}
\definecolor{asmita_color}{RGB}{198, 0, 0}
\definecolor{rajesh_color}{rgb}{0.93, 0.53, 0.18}

\newcommand{\rajesh}[1]{{\color{rajesh_color} {#1}}}

\begin{document}

\title{Azimuthal asymmetries in $J/\psi$-photon production at the EIC }
 \author{Dipankar Chakrabarti}
 \email{dipankar@iitk.ac.in}
 \affiliation{Department of Physics, Indian Institute of Technology Kanpur, Kanpur-208016, India}
 
 \author{Raj Kishore}
 \email{kishore@iitk.ac.in}
 \affiliation{Department of Physics, Indian Institute of Technology Kanpur, Kanpur-208016, India}

 \author{Asmita Mukherjee}
 \email{asmita@phy.iitb.ac.in}
 \affiliation{Department of Physics, Indian Institute of Technology Bombay, Mumbai-400076, India}

\author{Sangem Rajesh}
 \email{sangem.rajesh@vit.ac.in}
 \affiliation{Department of Physics, School of Advanced Sciences, Vellore Institute of Technology, Vellore,
Tamil Nadu 632014, India}
 \affiliation{INFN, sezione di Perugia, via A. Pascoli snc, 06123, Perugia, Italy}

\date{\today}
\begin{abstract}
We calculate azimuthal asymmetries  in back-to-back production of $J/\psi$ and a photon  in electron-proton scattering process at the future electron-ion collider (EIC) using TMD factorization framework. We consider the cases where the proton is  unpolarized or transversely polarized. For the formation of $J/\psi$, non-relativistic QCD (NRQCD) is used. We find that the cross-section gets contribution from only one color octet state, as a result, the azimuthal modulations become independent of the  long-distance matrix elements (LDMEs), and thus can be used to probe, in particular the gluon TMDs which are dominant in this kinematics. We show estimates of the upper bounds of different azimuthal asymmetries using the positivity bounds on TMDs. In addition, we show estimate of asymmetries using  the Gaussian parametrization of the TMDs. 
\end{abstract}
\maketitle

\section{Introduction}\label{sec1}

The quest for a three-dimensional representation of the nucleons in the momentum space is fundamentally dependent on transverse momentum dependent parton distributions (TMDs) \cite{Mulders:1995dh,boer1998time,boer2000angular,anselmino1999phenomenology,anselmino1995single,barone2002transverse}, which are  nonperturbative objects. These carry information on the inherent transverse motion of the partons and the correlations between the transverse momenta and the spins of the parent nucleons and partons. TMDs can be thought of as an extension of one-dimensional, collinear parton distribution functions (PDFs). Over a few decades, study of the TMDs has been of great interest to the scientific community in hadron physics. Both  experimental and theoretical studies on the semi-inclusive deep-inelastic scattering (SIDIS) processes have brought a fairly significant information on the quark TMDs but very less we have achieved yet, on the gluon TMDs. TMDs, unlike PDFs, are not universal in general. The operator structure of the quark TMDs  and the gluon TMDs, involve gauge links or Wilson lines, in order to make them gauge invariant quantities \cite{collins2002leading,ji2002parton,belitsky2003final,
boer2003universality}. The presence of the Wilson line introduces its sensitivity to the color flows and soft gluon exchanges in the processes in which they are probed. Hence, they are process dependent objects. Moreover, the process dependence of the gluon TMDs are more involved due to the presence of two gauge links than the quark TMDs which include one \cite{buffing2013generalized}. One such example has been predicted with the Sivers functions, basically the Sivers function observed in a SIDIS process is expected to have opposite sign as compared with its observation in Drell-Yan (DY) process, this is due to the difference in the gauge link structure, or in other words, initial/final state effects \cite{boer2003universality}. Recent data from RHIC are in favour of this theoretical prediction\cite{Anselmino:2016uie}, however, more data is needed to make a firm understanding of the process dependence of the TMDs. Experimental verification of such properties could test our understanding of the TMD formalism and nonperturbative QCD effects in general.     

At leading-twist, there are eight gluon TMDs \cite{mulders2001transverse}. Among them, linearly polarized gluon distribution \cite{mulders2001transverse} inside an unpolarized proton and the gluon Sivers function \cite{sivers1990,sivers1991} inside a transversely polarized proton have attracted a lot of interest in the last few years. Linearly polarized gluon TMD, also referred to as azimuthal correlated gluon distribution  causes azimuthal asymmetries, like $\text{cos}2\phi$ and affects the transverse momentum distribution of the outgoing particles in unpolarized electron-proton scattering process \cite{Pisano:2013cya}. On the other hand the gluon Sivers function, which describes correlation between intrinsic transverse motion of the unpolarized gluon and the transverse spin of the parent hadron, could also cause azimuthal asymmetries and single spin asymmetries \cite{sivers1990,sivers1991} in scattering processes, like electron-proton collision, where the proton is transversely polarized with respect to its direction of motion.        

In general, we do not have much information on the gluon TMDs yet, except the theoretical positivity bounds \cite{mulders2001transverse, Bacchetta:2020gko,Bacchetta:2017gcc}. They could play a crucial role in two-scale processes, such as SIDIS, where one measures the transverse momentum of the outgoing particle in addition to the photon virtuality, and the DY process, where the transverse momentum of the outgoing lepton-pair provides the second scale. The generalized factorization using TMDs can be used for these kinds of processes. However, TMD factorization has not been proven for all processes \cite{Echevarria:2019ynx}. Many proposals have been put forward in order to probe the gluon TMDs both in the lepton-proton and proton-proton collisions processes \cite{Marquet:2017xwy,Pisano:2013cya,boer2009dijet,efremov2018measure,efremov2018ratio,lansberg2017associated,dumitru2019measuring,sun2011gluon,boer2013determining,boer2012linearly,Echevarria:2015uaa,boer2012polarized,mukherjee2017linearly,mukherjee2016probing,rajesh2018sivers,kishore2020sivers,umberto2019jpsijet,DAlesio:2020eqo,DAlesio:2021yws}, however, experimentally, they are harder to be extracted as compared with the quark TMDs. Most of the proposals are based on analyzing the transverse momentum dependent azimuthal asymmetries in the heavy quark-antiquark pair productions or the bound state quarkonia productions because they are sensitive to the gluon content of hadrons, specially in the small longitudinal momentum fraction, $x$, domain\cite{rajesh2018sivers}. Depending on the types of gauge links present in the operator structure of the gluon TMDs \cite{buffing2013generalized}, which corresponds to the configurations, namely, one past and one future pointing gauge links, $[+,-]$ or $[-,+]$ and  both past or both future-pointing gauge links, $[-,-]$ or $[+,+]$, they are called f-type and d-type, respectively, each having a different operator structure \cite{buffing2013generalized}. In the small-$x$ domain, they are described in the literature as dipole \cite{dominguez2012linearly} and Weizsacker-Williams (WW)\cite{kovchegov1998gluon,mclerran1999fock} gluon distributions, respectively. 

In this work, we present a study of azimuthal asymmetries in  $J/\psi$ -photon production in the electron-proton scattering process: $ep\to eJ/\psi \gamma X$ \cite{Kim:1992at,Cacciari:1996zu}, where we consider both the unpolarized and transversely polarized initial protons.  The associated production of an energetic photon besides $J/\psi$ allows the kinematics where, their individual transverse momenta may not be constrained to be small, in this case one can  assume the TMD factorization for this process as the soft scale is provided by the total transverse momentum of the outgoing particles, $J/\psi$+ photon, which is required to be much smaller than its invariant mass. This kinematic condition can be achieved by considering a channel where $J/\psi$ and photon are almost back-to-back in the transverse plane. This scenario provides an advantage that one can access the TMDs over a range of scales by varying the invariant mass of the pair. This  process probes the kinematic region of $z<1$, where $z$ is the fraction of energy of the virtual photon taken by $J/\psi$ in the proton rest frame. 

For the calculation of $J/\psi$ production rate, we have employed the non-relativistic QCD (NRQCD) effective field theory framework \cite{hagler2001towards,yuan2001polarizations,yuan2008heavy}, which is extensively used since it gives a rigorous description of heavy quarkonium generation and decay\cite{Bodwin:1994jh,bodwin1995rigorous}. 
Within this framework, the amplitude for a quarkonium production, here a $J/\psi$,  can be given as factorization of the hard process, that corresponds to production of $c\bar{c}$ pair and the soft process of hadronization of $c\bar{c}$ pair to $J/\psi$ state. A nonperturbative quantity called long-distance matrix elements (LDMEs) encodes the soft hadronization process which describes the transition probability to form a quarkonium state from the heavy quark pair that are produced in the hard process \cite{Bodwin:1994jh}. This is expected to be process independent and can be extracted by fitting data. Several LDME sets exist in the literature. In terms of the heavy-quark velocity parameter $v$, LDMEs are anticipated to scale with a specific power of $v$, in the limit $v<<1$ \cite{Boer:2021ehu,lepage1992improved}. With $v^2\approx 0.3$ for charmonium, NRQCD introduces an expansion in $v$ in the cross section, besides the common expansion in running coupling constant $\alpha_s$. As a result, the heavy quarks pair could produce in different quantum states that can be represented as $^{2S+1}L_J^{(1,8)}$. Here, $S$, $L$ and $J$ represent their spin, orbital and total angular momentum, respectively. The color configuration is given by $(1,8)$, where $(1)$ represents color singlet (CS) and $(8)$ represents color octet (CO), i.e, NRQCD incorporates both the CS and CO contributions to the cross section, in general. 

In our recent publications, we studied the azimuthal asymmetries in $J/\psi$ \cite{kishore2021cos2phi} and $J/\psi+$ jet \cite{Kishore:2019fzb,kishore2022cos2phi} productions in the electron-proton collision, where we probed the similar kinematics of $z<1$ at the next-to-leading order in $\alpha_s$. We considered the TMD formalism and $J/\psi$ production rate has been calculated using NRQCD framework. In many cases, azimuthal asymmetries that are suggested to probe gluon TMDs  show a significant dependency on the choice of the LDME set \cite{kishore2021cos2phi,kishore2022cos2phi}. This introduces  some uncertainty in the extraction of the gluon TMDs from these observables. In the process considered here, as we have shown below, the azimuthal asymmetries are independent of the LDMEs. This is because only one color octet state $^3S_1^{(8)}$ contributes at the cross-section level. We present some numerical estimates of the upper bounds of the asymmetries using model independent positivity bounds on the TMDs \cite{mulders2001transverse} as well as using the Gaussian parametrization.      

The paper is organized as follows. Introduction of the paper is given in Sec.\ref{sec1}. The theoretical formalism and azimuthal asymmetries are presented respectively in Sec.\ref{sec2} and \ref{sec3}. Sec.\ref{sec4} and \ref{sec5} discuss the numerical results and conclusion of the paper. Analytic results for the amplitude modulations are given in the appendix.

\section{Formalism }\label{sec2}
We consider the production of associated $J/\psi+\gamma$ in  (un)polarized electron-proton collision process 
\be
e(l)+p^\uparrow(P)\rightarrow e(l^\prime)+ J/\psi (P_\psi)+\gamma (p_\gamma)+X\,,
\ee
where, the 4-momenta of each particle is given in the round brackets, and the transverse polarization of the proton is represented with an arrow in the superscript. We consider the photon-proton center of mass frame, wherein the photon and proton move along the $z$-axis. The 4-momenta of the target  proton $P$ and the virtual photon $q$ are given by
\be\label{pq4m}
P^\mu&=&n^\mu_-+\frac{M_p^2}{2}n^\mu_+\approx n^\mu_-\,,\nonumber\\
q^\mu&=&-\xB n^\mu_-+\frac{Q^2}{2\xB}n^\mu_+\approx -\xB P^\mu+(P\cdot q)n^\mu_+\,,
\ee
where, the $n_+$ and $n_-$ are two light-like vectors with $n_+^2=n_-^2=0$ and $n_+\cdot n_-=1$. The invariant mass of the virtual photon is $Q^2=-q^2$ and the Bjorken variable, $\xB=\frac{Q^2}{2P\cdot q}$. Mass of the proton is denoted by  $M_p$. The center-of-mass (cm) energy of the electron-proton  system is $S=(P+l)^2=2P\cdot l=\frac{2P\cdot q}{y}$ and that leads to having $Q^2=\xB y S$, here $y=\frac{P\cdot q}{P\cdot l}$ is the energy fraction carried by the photon. The virtual photon-proton invariant mass is defined as $W^2_{\gamma p}=(q+P)^2=\frac{Q^2(1-\xB )}{\xB }=yS-Q^2$.  The 4-momentum of the incoming lepton reads
\be\label{q4m2}
l^\mu&=&\frac{1-y}{y}\xB n^\mu_-+\frac{1}{y}\frac{Q^2}{2\xB }n^\mu_++\frac{\sqrt{1-y}}{y}Q\hat{l}^\mu_\perp\,,
\ee
here, $\hat{l}_\perp^\mu$ is the unit transverse vector.\par
\begin{figure}[H]
\begin{center} 
\includegraphics[height=5cm,width=6cm]{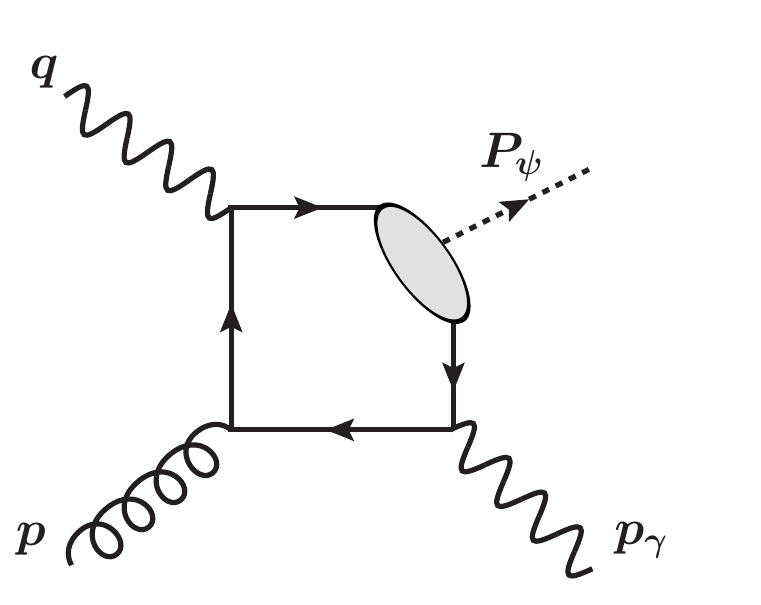}
\end{center}
\caption{\label{fig_feyn} Feynman diagram representation of $J/\psi$-photon production in SIDIS process.}
\end{figure}

The differential cross section within the TMD  factorization framework can be written as \cite{Pisano:2013cya}
\begin{equation}\label{2d2}
 \begin{aligned}
 \d\sigma={}&\frac{1}{2S}\frac{\d^3{\bm l^\prime}}{(2\pi)^32E_{l^\prime}}
\frac{\d^3{\bm P}_\psi}{(2\pi)^32E_\psi}\frac{\d^3{\bm p}_\gamma}{(2\pi)^32E_\gamma}\int \d x \,\d^2 {\bm p}_\sT \, (2\pi)^4 \,\delta^4(q+p-P_\psi-p_\gamma)\\
& \times\frac{1}{Q^4}L^{\mu\nu}(l,q)\,\Phi^{\rho\sigma}_g(x,{\bm p_\sT})\, H_{\mu\rho}H_{\nu\sigma}^\ast\,.
\end{aligned}
\end{equation}
 In the above equation, the leptonic tensor, $L^{\mu\nu}$, has the standard form
\begin{eqnarray}\label{lep:ep}
L^{\mu\nu}&=&e^2 Q^2\left(-g^{\mu\nu}+\frac{2}{Q^2}(l^\mu l^{\prime\nu}+l^\nu l^{\prime\mu})\right)\,,
\label{eq:lep1}
\end{eqnarray}
where the averaging over spins of the initial lepton is assumed, and the 4-momentum of the final scattered lepton is $l^\prime=l-q$. Using Eqs.\eqref{pq4m} and \eqref{q4m2}, the leptonic tensor can be recast in the following form
\begin{equation}\label{lep:ep1}
 \begin{aligned}
L^{\mu\nu}={}&e^2 \frac{Q^2}{y^2}\Big[-(1+(1-y)^2)g_T^{\mu\nu}+4(1-y)\epsilon_{ L}^\mu \epsilon_{ L}^\nu+4(1-y)\left(\hat{l}^\mu_\perp \hat{l}^\nu_\perp +\frac{1}{2} g_T^{\mu\nu} \right)\\
&+2(2-y)\sqrt{1-y}\left(\epsilon_{ L}^\mu \hat{l}^\nu_\perp + \epsilon_{ L}^\nu \hat{l}^\mu_\perp \right)\Big]\,,
\end{aligned}
\end{equation}
where the transverse metric tensor is defined as $g_{T}^{\mu\nu}=g^{\mu\nu}-n_+^{\mu}n_-^{\nu}-n_+^{\nu}n_-^{\mu}$, and ~
\be
\epsilon_{ L}^\mu(q) =\frac{1}{Q}\left(q^\mu +\frac{Q^2}{P\cdot q} P^\mu\right)\,,
\ee 
is the longitudinal polarization vector of the virtual photon with $\epsilon_{ L}^{2}(q)=1$ and $\epsilon_{ L}^\mu(q) q_\mu=0$. The $H$ contains the scattering amplitude of virtual photon-gluon fusion:~$\gamma^\ast(q) +g(p) \rightarrow J/\psi(P_\psi) + \gamma(p_\gamma)\,$ partonic process whose corresponding Feynman diagram is shown in Fig.\ref{fig_feyn}. Moreover, at leading order one more partonic process:~ $\gamma^\ast(q)+\gamma(p)\to J/\psi(P_\psi)+\gamma(p_\gamma)$ can contribute to our process, wherein the emission of photon off the proton  can happen elastically or inelastically. In this process, $J/\psi$ production happens through color singlet mechanism. However, its  contribution is insignificant as compared with the virtual photon-gluon fusion subprocess due to the much higher density of gluons than photons in the proton; as discussed in Ref. \cite{Kniehl:2006qq}. Although the CO states, for example $\langle 0 \vert {\cal O} (\leftidx{^{3}}{S}{_1}^{(8)})\vert 0 \rangle $, are suppressed relative to $\langle 0 \vert {\cal O} (\leftidx{^{3}}{S}{_1}^{(1)})\vert 0 \rangle $ by $O(v^4)$, this does not overcome the number density suppression.

We use  the NRQCD framework for $J/\psi$ production \cite{baier1983hadronic,boer2012polarized}, where at leading order, only the CO state $\leftidx{^{3}}{S}{_1}^{(8)}$  contributes to the $J/\psi$ production in the partonic process:$\gamma^\ast(q) +g(p) \rightarrow J/\psi(P_\psi) + \gamma(p_\gamma)\,$, hence the cross section contains only one LDME i.e.,  $\langle 0 \vert {\cal O} (\leftidx{^{3}}{S}{_1}^{(8)})\vert 0 \rangle $. As a result, the asymmetry becomes independent of the choice of LDMEs, and can be used to extract the gluon TMDs.
The 4-momenta of the gluon,  $J/\psi$  and final photon are the followings
\be
p^\mu&\backsimeq&x P^\mu +p^\mu_T \,,\nonumber\\
P_\psi^\mu&=&z(P\cdot q)n^\mu_+ + \frac{M_\psi^2+{\bm P}_{\psi \perp}^2}{2zP \cdot q}P^\mu+P^\mu_{\psi \perp}\,,\nonumber\\
p_\gamma^{ \mu}&=&(1-z)(P\cdot q)n^\mu_+ + \frac{{\bm p}_{\gamma \perp}^{ 2}}{2(1-z)P \cdot q}P^\mu+p^{\mu}_{\gamma \perp}\,,
\ee
where $x=p\cdot n_+$ and $p_T$ are respectively the longitudinal momentum fraction and transverse momentum of the  gluon. The inelastic variable $z=\frac{P\cdot P_\psi}{P\cdot q}$, is defined as the energy fraction of the virtual photon carried by $J/\psi$ in the proton rest frame. The $P_{\psi\perp}$ and $p_{\gamma\perp}$ are transverse momenta of the $J/\psi$ and photon respectively. $M_\psi$ represents the quarkonium mass.\par
 At the partonic level, the Mandelstam variables are defined as
 \begin{eqnarray}
 \hat{s}&=&(p+q)^2=Q^2\left(\frac{x-\xB}{\xB}\right)\,,\nonumber\\
 \hat{t}&=&(q-P_\psi)^2=\frac{1}{z}(z-1)\left(zQ^2+M^2_\psi\right)-\frac{\bm P^2_{\psi \perp}}{z}\nonumber\,,\\
 \hat{u}&=&(p-P_\psi)^2=M^2_\psi-x z \frac{Q^2}{\xB}\,.
\end{eqnarray}
In Eq.\eqref{2d2}, the gluon correlator, $\Phi_g^{\mu\nu}$, a nonperturbative quantity, contains the dynamics of gluons inside a proton. For an unpolarized proton, its parametrization in terms of the gluon TMDs is given by \cite{mulders2001transverse,meissner2007,Boer_2016}

\be\label{gc:un}
\Phi_{U}^{\mu\nu}(x,{\bm p}_\sT)=\frac{1}{2x}\Bigg\{-g_{T}^{\mu\nu}f_1^g(x,{\bm p}^2_\sT)+\left(\frac{p_{\sT}^{\mu} p_{\sT}^{\nu}}{M_p^2}+g_{T}^{\mu\nu}\frac{{\bm 
p}^2_\sT}{2M_p^2}\right)h^{\perp g}_1(x,{\bm p}^2_\sT)\Bigg\}\,,
\ee
where the $f_1^g$ and $h_1^{\perp g}$, T-even TMDs, encode the distribution of unpolarized and linearly polarized gluons respectively. These TMDs can be non-zero, even if, initial and final state interactions are absent in the process. Similarly, for the transversely polarized proton, with their transverse spin vector $\bm{S}_\sT$, we have \cite{mulders2001transverse,meissner2007,Boer_2016}
\be
\begin{aligned}\label{gc:T}
\Phi_T^{\mu\nu}(x,\bm p_\sT )={}& 
\frac{1}{2x}\bigg \{-g^{\mu\nu}_T
    \frac{ \epsilon^{\rho\sigma}_T p_{\sT \rho} S_{T\sigma}}{M_p}
f_{1T}^{\perp\,g}(x, \bm p_\sT^2) + i \epsilon_T^{\mu\nu}
    \frac{ p_\sT \cdot  S_T}{M_p} g_{1T}^{g}(x, \bm p_\sT^2) \\
&+  
\frac{p_{\sT \rho}\epsilon_{T}^{\rho\{ \mu}p_{\sT}^{\nu\}}} 
{2M_p^2} \frac{p_{\sT }\cdot S_T}{M_p} h_{1T}^{\perp g}(x, \bm p_\sT^2)
-\frac{p_{\sT \rho}\epsilon_T^{\rho \{\mu}S_T^{\nu\}}+ 
S_{T\rho}\epsilon_T^{\rho\{\mu}p_\sT^{\nu\}}}{4M_p}h_{1T}^{g}(x, \bm p_\sT^2)  
 \bigg \}\,,
\end{aligned}
\ee
where the notations are: antisymmetric tensor $\epsilon_T^{\mu\nu}=\epsilon^{\mu\nu\rho\sigma} P_\rho n_{+\sigma}$ with $\epsilon_T^{12}=+1$ and the symmetric tensor $p_{\sT \rho}\epsilon_{T}^{\rho\{ \mu}p_{\sT}^{\nu\}}=p_{\sT \rho}(\epsilon_T^{\rho\mu}p_{\sT}^\nu+\epsilon_T^{\rho\nu}p_{\sT }^\mu)$. In Eq.\eqref{gc:T},  we have three T-odd TMDs: the Sivers function, $f_{1T}^{\perp g}$,  describes the density of unpolarized gluons, while $h_{1T}^{\perp g}$ and $h_{1T}^g$, are  linearly polarized gluon densities  of a transversely polarized proton. The $g_{1T}^g$, T-even TMD,  is the distribution of circularly polarized gluons in a transversely polarized proton, which does not contribute when the lepton is unpolarized, as it is in the antisymmetric part of the correlator. \par
The momentum conservation delta function, given in  Eq.\eqref{2d2},  can be decomposed as follows,
\begin{equation}\label{dfun}
\begin{aligned}
 \delta^{4}\bigl(q+p-P_{\psi}-p_{\gamma }\bigr)
 & =\frac{2}{yS}\delta\bigl(1-z-\bar{z}\bigr)\delta\left(x-\frac{\bar{z}(M^2_\psi+ \bm{P}_{\psi\perp}^2)+z\bm{p}^{ 2}_{\gamma\perp} + z \bar{z}Q^2}{z(1-z)yS}\right )\delta^{2}\bigl(\bm{p}_{\sT}- \bm{P}_{\psi\perp}- \bm{p}_{\gamma\perp}\bigr)\;,
\end{aligned}
\end{equation}
here $\bar{z}=\frac{P.p_\gamma}{P.q}$ is the energy fraction carried by the final photon. 
 The phase-space of outgoing particles  is given by
\be\label{felectron}
 \frac{\d^3{\bm l^\prime}}{(2\pi)^32E_{l^\prime}}=\frac{1}{16\pi^2}\d Q^2\d y\,,\quad
\frac{\d^3{\bm P}_\psi}{(2\pi)^32E_\psi}=\frac{\d^2{\bm P}_{\psi \perp}\d z}{(2\pi)^32z}\,,\quad
\frac{\d^3{\bm p}_\gamma}{(2\pi)^32E_\gamma}=\frac{\d^2{\bm p}_{\gamma \perp}\d \bar{z}}{(2\pi)^32\bar{z}}\,.
\ee
After integrating over $\bar{z}$, $\bm{p}_{_T}$ and $x$ using Eq.\eqref{dfun}, the cross section in Eq.\eqref{2d2} can be rewritten as
\begin{equation}
\begin{aligned}\frac{\mathrm{d}\sigma}{\mathrm{d}z\,\mathrm{d}y\,\mathrm{d}Q^2 \,\mathrm{d}^{2}\bm{q}_{\sT}\mathrm{d}^{2}\bm{K}_{\perp}} & =\frac{1}{(2\pi)^{4}}\frac{1}{16yS^2z(1-z)Q^{4}}L^{\mu\nu}(\ell,q)\,  \Phi^{\rho\sigma}_{g}(x,\bm{q}_{\sT})\, H _{\mu\rho}\, H^*_{\nu\sigma}\,.
\end{aligned}
\label{eq:csfinal}
\end{equation}
In the above equation, we have introduced the following transverse vectors
\begin{equation}
\begin{aligned}
\bm{q}_{\sT} & \equiv  \bm{P}_{\psi\perp}+ \bm{p}_{\gamma\perp}\,,\quad \bm{K}_{\perp}\equiv\frac{ \bm{P}_{\psi\perp}- \bm{p}_{\gamma\perp}}{2}\,.
\end{aligned}
\end{equation}
\begin{figure}[H]
\begin{center} 
\includegraphics[height=5cm,width=10cm]{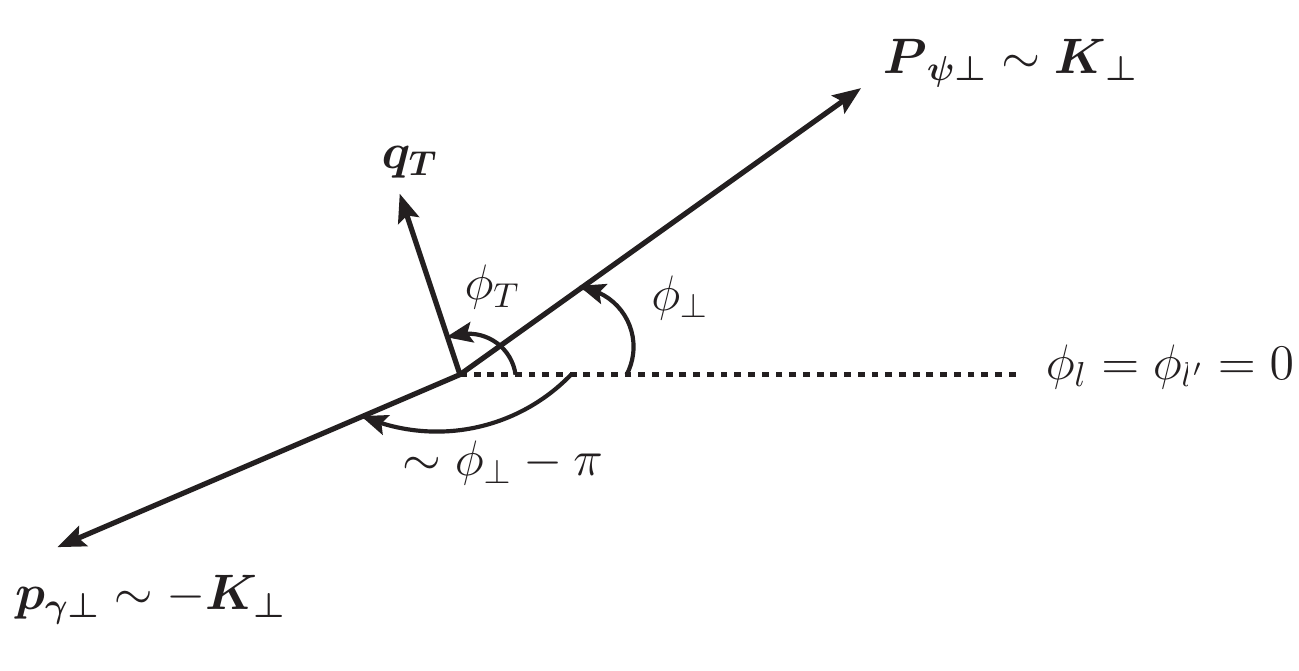}
\end{center}
\caption{\label{fig_angles} Representation of azimuthal angles in $J/\psi$-photon production in SIDIS process.}
\end{figure}
\section{Azimuthal Asymmetries }\label{sec3}

We consider the kinematical region; $q_\sT\ll K_\perp$, where $q_\sT\equiv\vert\bm{q}_\sT\vert$ and $K_\perp\equiv\vert\bm{K}_\perp\vert$, which allows assuming TMD factorization for the considered process. This kinematical condition leads to the scenario where the outgoing particles, particularly $J/\psi$ and photon, are almost back-to-back in the transverse plane w.r.t. the line of collision of virtual photon-proton system as illustrated in Fig.~\ref{fig_angles}. We can then approximate that $\bm{P}_{\psi\perp}\simeq -\bm{p}_{\gamma\perp}\simeq\bm{K}_\perp$. With these kinematics, one can write the differential cross-section as \cite{Pisano:2013cya}
\begin{equation}
\frac{\d\sigma}
{\d z\,\d y\,\d\xB \,\d^2\bm{q}_{\sT} \d^2\bm{K}_{\perp}} \equiv \d\sigma (\phi_S, \phi_\sT,\phi_\perp) =    \d\sigma^U(\phi_\sT,\phi_\perp)  +  \d\sigma^T (\phi_S, \phi_\sT,\phi_\perp)  \,.
\label{eq:cs}
\end{equation}
where, for the case of unpolarized proton,
\begin{align}\label{eq:Un}
\d\sigma^{U} & =\mathcal{N}\bigg[\bigl(\mathcal{A}_{0}+\mathcal{A}_{1} \cos\phi_{\perp}+\mathcal{A}_{2} \cos2\phi_{\perp}\bigr)f_{1}^{g}(x,\bm{q}_{\sT}^{2})+\bigl(\mathcal{B}_{0}  \cos2\phi_{\sT}+\mathcal{B}_{1}  \cos(2\phi_{\sT}-\phi_{\perp})\nonumber\\
 & \qquad\quad+\mathcal{B}_{2}  \cos2(\phi_{\sT}-\phi_{\perp})+\mathcal{B}_{3}  \cos(2\phi_{\sT}-3\phi_{\perp})+\mathcal{B}_{4}  \cos(2\phi_{\sT}-4\phi_{\perp})\bigr)\frac{ \bm q_\sT^2  }{M_p^2} \,h_{1}^{\perp\, g} (x,\bm{q}_{\sT}^2)\bigg]\,,
\end{align}
while, for the case of transversely polarized proton,
\begin{align}\label{eq:Tr}
\d\sigma^T
  & = \mathcal{N} {\vert \bm S_\sT\vert}\, \bigg[\sin(\phi_S -\phi_\sT) \bigl( \mathcal{A} _0 + \mathcal{A} _1 \cos \phi_\perp  + \mathcal{A} _2 \cos 2 \phi_\perp \bigr ) \frac{ |\bm q_\sT | }{M_p}f_{1T}^{\perp\, g} (x,\bm{q}_{\sT}^2)\nonumber \\
 & + \cos(\phi_S-\phi_\sT) \bigl( \mathcal{B} _0 \sin 2 \phi_\sT  +  \mathcal{B} _1 \sin(2\phi_\sT-\phi_\perp) + \mathcal{B} _2 \sin 2 (\phi_\sT-\phi_\perp) \nonumber \\
 &\quad+ \mathcal{B} _3 \sin( 2\phi_\sT-3\phi_\perp)  + \mathcal{B} _4\sin(2\phi_\sT- 4\phi_\perp)  \bigr) \frac{ |\bm q_\sT |^3  }{M_p^3} \,h_{1T}^{\perp\, g} (x,\bm{q}_{\sT}^2) \nonumber\\
& + \bigl (\mathcal{B}_0  \sin(\phi_S+\phi_\sT) + \mathcal{B}_1  \sin(\phi_S + \phi_T-\phi_\perp) + \mathcal{B}_2  \sin(\phi_S+\phi_\sT-2\phi_\perp) \nonumber \\
&  \quad  +  \mathcal{B}_3  \sin(\phi_S+\phi_\sT-3\phi_\perp) + \mathcal{B}_4  \sin(\phi_S+\phi_\sT-4 \phi_\perp) \bigr)\frac{ |\bm q_\sT | }{M_p} h_{1T}^{g} (x,\bm{q}_{\sT}^2) \bigg ]\, ,
\end{align}
where $\mathcal{N}=\frac{1}{(2\pi)^{4}}\frac{1}{16yS^2z(1-z)Q^{4}}$ is the kinematical factor.   $\phi_S$, $\phi_\sT$ and $\phi_\perp$  are  respectively  azimuthal angles of the three-vectors $\bm S_\sT$,  $\bm q_\sT$ and  $\bm K_\perp$ that are measured   w.r.t. the lepton plane ($\phi_{\ell}=\phi_{\ell^\prime}=0)$ as shown in Fig.~\ref{fig_angles}. The amplitude modulations $\mathcal{A}_i$ with $i=0,1,2$ and $\mathcal{B}_j$ with $j=0,1,2,3,4$  are amplitude square of the $\gamma^\ast +g \rightarrow J/\psi + \gamma$ process, which are given in the appendix.\ref{appen}.\par
The cross section in Eq.\eqref{eq:Un} and Eq.\eqref{eq:Tr} has azimuthal modulations that allow\rajesh{s} to extract the specific gluon TMD by measuring the weighted azimuthal asymmetry.\par
The weighted azimuthal asymmetry, gives the ratio of specific gluon TMD over unpolarized $f_1^g$ and  is defined as \cite{DAlesio:2019qpk}
\begin{align}
A^{W(\phi_S,\phi_\sT)} & \equiv 2\,\frac {\int\d \phi_S\, \d \phi_\sT \,\d\phi_\perp\, W(\phi_S,\phi_\sT)\,\d\sigma(\phi_S,\,\phi_\sT,\,\phi_\perp)}{\int \d \phi_S\,\d\phi_\sT \,\d\phi_\perp\,\d\sigma(\phi_S,\phi_\sT,\phi_\perp)} \,,
\label{eq:mom}
\end{align}
where $W(\phi_S,\phi_\sT)$ is the weight factor and  the denominator is given by
\begin{align}\label{eq:f1}
\int \d\phi_S\,\d \phi_\sT \,\d\phi_\perp\,\d\sigma(\phi_S,\phi_\sT,\phi_\perp)  & =\int \d\phi_S\,\d \phi_\sT \,\d\phi_\perp\,\d\sigma^U(\phi_\sT,\phi_\perp)=  (2\pi)^3 \mathcal{N}  {\cal A}_0 f_{1}^{g}(x,\bm{q}_{\sT}^{2})\;.
\end{align}
By integrating over the azimuthal angle $\phi_\perp$, the transversely polarized cross-section, Eq.~(\ref{eq:Tr}), can be simplified further as
\begin{align}\label{eq:sivers}
\int\d\phi_\perp\d\sigma^T & =  2\pi\mathcal{N} {\vert \bm S_\sT\vert}\, \frac{ |\bm q_\sT | }{M_p} \left [\mathcal{A} _0 \sin(\phi_S -\phi_\sT)    f_{1T}^{\perp\, g} (x,\bm{q}_{\sT}^2) -\frac{1}{2} \mathcal{B} _0\sin(\phi_S-3\phi_\sT)   \frac{ |\bm q_\sT |^2 }{M_p^2} \,h_{1T}^{\perp\, g} (x,\bm{q}_{\sT}^2) \right . \nonumber\\*
  &\qquad+\mathcal{B}_0  \sin(\phi_S+\phi_\sT)  h_{1}^{g} (x,\bm{q}_{\sT}^2) \bigg ]\,,
\end{align}
where we have used the relation
\begin{equation}
h_1^g \equiv h_{1T}^g +\frac{\bm p_\sT^2}{2 M_p^2}\,  h_{1T}^{\perp\,g},
\label{eq:h1}
\end{equation}
which vanishes in the collinear configuration \cite{Boer_2016dijet} unlike their counterpart, quark transversity TMD.
The  $h_1^{\perp\, g}$ gluon TMD could be extracted by studying the following two azimuthal asymmetries, for specific weight factors : 
\begin{align}
A^{\cos 2 \phi_\sT}&= \frac{\bm q_\sT^2}{M_p^2} \, \frac{ {\cal B}_0  }{{\cal A}_0}\,\frac{h_1^{\perp\, g}(x,\bm q_\sT^2 )}{ f_1^{g}(x,\bm q_\sT^2 )}\,,
\label{eq:cos2phiT}
\end{align}
and
 \begin{align}
A^{\cos 2 (\phi_\sT-\phi_\perp)}&= \frac{\bm q_\sT^2}{M_p^2} \, \frac{{\cal B}_2  }{{\cal A}_0}\,\frac{h_1^{\perp\, g}(x,\bm q_\sT^2 )}{ f_1^{g}(x,\bm q_\sT^2 )}\,.
\label{eq:cos2phiT2phiP}
\end{align}
Using Eq.\eqref{eq:sivers} with ${\vert \bm S_\sT\vert}=1$ , one could exploit the asymmetries given below to extract the $f_{1T}^{\perp g}$, $h_1^g$ and $h_{1T}^{\perp g}$ TMDs
\begin{align}\label{eq:f1t_sivers}
A^{\sin(\phi_S-\phi_\sT)} & =  \frac{\vert \bm q_\sT\vert}{M_p}\, \frac{ f_{1T}^{\perp\,g}(x,\bm q_\sT^2) }{f_1^g(x,\bm q_\sT^2)}\,,
\end{align}
\begin{align} \label{eq:h1g-asy}
A^{\sin(\phi_S+\phi_\sT)}  & =  \frac{\vert \bm q_\sT\vert}{M_p}\,\frac{{\cal B}_0 }{{\cal A}_0 } \frac{ h_{1 }^{g}(x,\bm q_\sT^2) }{f_1^g(x,\bm q_\sT^2)}\,,
 \end{align}
 and
\begin{align}\label{eq:h1Tg-asy}
A^{\sin(\phi_S-3\phi_\sT)}  & =   -  \frac{\vert \bm q_\sT\vert ^3}{2M_p^3}\, \frac{{\cal B}_0 }{{\cal A}_0 } \frac{ h_{1T}^{\perp\,g}(x,\bm q_\sT^2)}{f_1^g(x,\bm q_\sT^2)} \,.
\end{align}
In the Sivers asymmetry, $A^{\sin(\phi_S-\phi_\sT)}$, the amplitude modulation term $\mathcal{A} _0$ has been cancelled in the numerator and denominator, because we have fixed the kinematical variables  in the cross section. 

\section{Results}\label{sec4}

\subsection{Unpolarized Cross Section}

In this section, we present numerical results, first we start with unpolarized cross section. The free parameters entering  the cross section are the long distance matrix elements (LDMEs). There are several sets in the literature, although, they are expected to be universal. Different assumptions are adopted  in their extractions resulting several sets; for instance, theoretical accuracy of perturbative calculation, kinematical cuts imposed mainly on the transverse momentum and chosen data sets. As aforementioned, only the CO state $\leftidx{^{3}}{S}{_1}^{(8)}$ contributes to the present study. We consider  $\langle 0\mid \mathcal{O}^{J/\psi}(\leftidx{^{3}}{S}{_1}^{[8]})\mid 
0\rangle=2.24\times 10^{-3}~\mathrm{GeV}^3$  from Ref.\cite{Butenschoen:2011yh}. One could use  a different LDME set, even in such case, our results will be rescaled as per the value of $\langle 0\mid \mathcal{O}^{J/\psi}(\leftidx{^{3}}{S}{_1}^{[8]})\mid 0\rangle$.  In Ref.\cite{Butenschoen:2011yh}, LDMEs were extracted by performing a global fit on $J/\psi$ data from $pp$ collision,  within the NRQCD framework at NLO, and a kinematical cut  $p_{\sT\psi}> 3$ GeV was imposed on the transverse momentum of the $J/\psi$ in the fit.  \par
As shown in Eq.\eqref{eq:f1}, only the $\mathcal{A}_0$ term and gluon TMD $f_1^g(x,\bm{q}_T^2)$  contribute to the unpolarized cross section after integrating over the azimuthal angles. The modulation term $\mathcal{A}_0$, given in  Eq.\eqref{A0:mod}, matches with the results given in Ref.\cite{Kniehl:2006qq}.  For the parametrization of unpolarized gluon TMD, we follow a Gaussian parametrization of TMDs (see Ref.~\cite{mauro2002dy})  given as \cite{boer2012linearly,boer2012polarized}
\begin{eqnarray}\label{eq:gauss_f1}
f_1^g(x,\bm{q}_\sT^2)=f_1^g(x,\mu)\frac{e^{-\bm{q}_\sT^2/\langle q_\sT^2\rangle}}{\pi\langle q_\sT^2\rangle}\,,
\end{eqnarray}
where $f_1^g(x,\mu)$ is the collinear gluon PDF at the probing scale $\mu=\sqrt{M^2_\psi+Q^2}$ \cite{KNIEHL2002337}. We use MSTW2008 set \cite{Martin:2009iq} for collinear PDF. The Gaussian parametrization of  TMDs with  Gaussian width  $\langle q_\sT^2\rangle=1~\mathrm{GeV}^2$ for gluons, describes the data reasonably well as reported in Ref.\cite{DAlesio:2019gnu}, so we use the same parametrization here.  \par
The future EIC detector \cite{EIC} is expected to reach large integrated luminosities at different $\sqrt{S}$. Therefore, we consider two \sout{c.m.s} \rajesh{cm} energies, namely, $\sqrt{S}=45$ GeV and $\sqrt{S}=140$ GeV. The cross-section for $J/\psi+\gamma$ production could also receive a contribution from  diffractive scattering off the proton via pomeron exchange process, however such process contributes at $z\approx 1$. As a result, we impose an upper cut $z<0.9$ to avoid such contribution to the cross section. This kinematical cut also prevents hitting infrared divergences, see the Eq.\eqref{eq:csfinal}. Furthermore, we impose a lower cut  $0.3<z$ to avoid contribution via the resolved-photon channel, which contributes in the low $z$ regions.
The virtuality of the photon is restricted as  $3 < Q^2 < 100~\mathrm{GeV}^2$, to exclude photoproduction. For lower energy, $\sqrt{S}=45$ GeV, where the cross section is relatively small, we consider the  interval $10 < W_{\gamma p} < 40$ GeV. While for $\sqrt{S}=140$ GeV, we choose $20 < W_{\gamma p} < 80$ GeV, where $W_{\gamma p}$ is the invariant mass of the photon-proton system. $q_\sT$, the sum of the transverse momenta of the final $J/\psi$ and $\gamma$, which is equal to the transverse momentum of the initial gluon, is integrated in the interval $0 < q_\sT < 1$ GeV, while, their average transverse momentum is considered to be $K_\perp > 1$. With these choices of interval, we are safely in the limit, $|{\bm q}_\sT| \ll |{\bm K}_\perp|$ which corresponds to a scenario of almost back-to-back $J/\psi-\gamma$ pair production.  \par
\begin{figure}[H]
	\begin{center} 
		\begin{subfigure}{0.49\textwidth}
		\includegraphics[height=7cm,width=8.5cm]{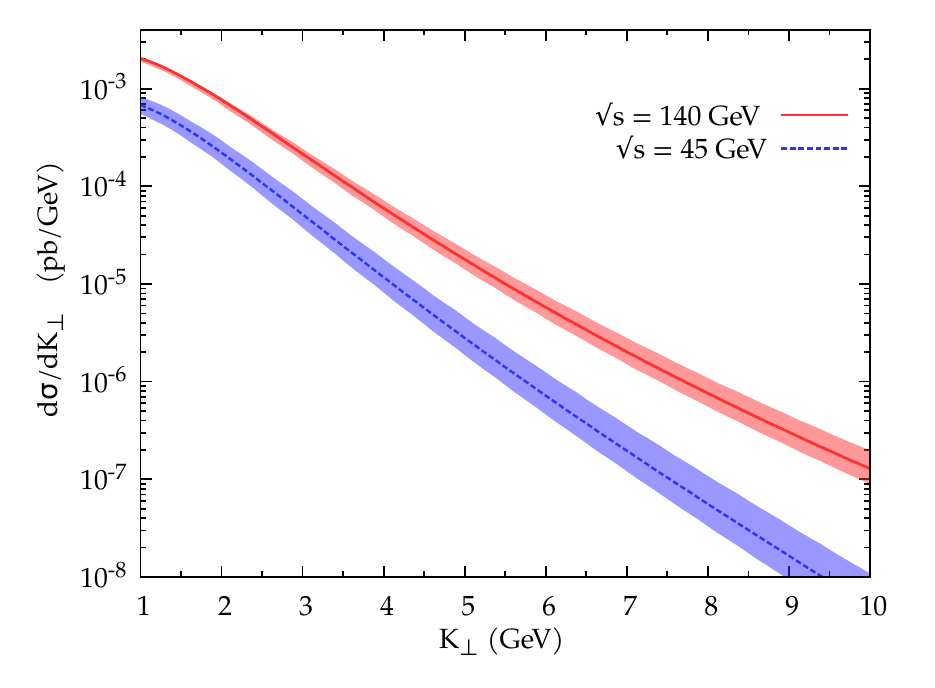}
			\caption{}
		\end{subfigure}
	    \begin{subfigure}{0.49\textwidth}
	    \includegraphics[height=7cm,width=8.5cm]{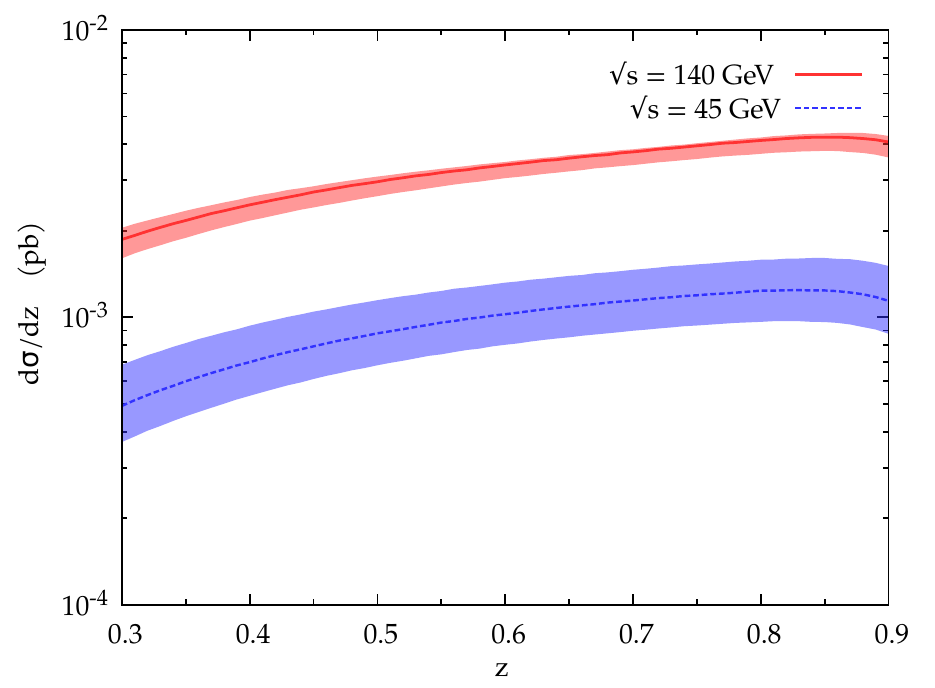}
	        \caption{}
	    \end{subfigure}
	\end{center}
	\caption{\label{fig:unc_scale} {Unpolarized differential cross section of $e+p\rightarrow e +J/\psi+\gamma +X$ process as a function of $K_\perp$ (left) and $z$ (right) at $\sqrt{S} = 45$ and 140 GeV. The kinematical cuts are $1 < K_{\perp} < 10$ GeV, $0 < q_\sT < 1$ GeV,  $0.3 < z < 0.9$. For  $\sqrt{s}= 140$ GeV we have taken $20 < W_{\gamma p} < 80$ while for $\sqrt{s}= 45$ GeV, $10 < W_{\gamma p} < 40$. The bands are obtained by varying the factorization scale in the range  $\frac12\mu < \mu  < 2\mu $. }}
\end{figure}

In Fig.\ref{fig:unc_scale}, we show the unpolarized cross section as a function of $K_\perp$ and $z$ for $\sqrt{S}=45$ and 140 GeV, where $z$ is the energy fraction carried by $J/\psi$. The theoretical uncertainties in Fig.\ref{fig:unc_scale} are obtained by varying the probing scale $[0.5\mu, 2\mu]$, and its variation is more evident in the lower energy  case compared to high energy. Comparing with $\sqrt{S}=140$  GeV,  the transverse momentum spectrum decreases very rapidly for $\sqrt{S}=45$  GeV, while the $z$-spectrum increases with $z$.  The  covered kinematical ranges of longitudinal momentum fraction are $x\approx [3\times 10^{-3}, 9\times 10^{-1}]$ and $x\approx [1\times 10^{-2}, 9\times 10^{-1}]$  at $\sqrt{S}=140$ GeV and  $45$ GeV respectively. The low $x$ range at $\sqrt{S}=140$ leads to high gluon density inside a proton, as a result the cross section is high compared to the low cm energy. Moreover, it leads to small theoretical uncertainty band in  Fig.~\ref{fig:unc_scale} for high cm energy.
\par
From Fig.~\ref{fig:unc_scale}, we obtain the integrated  cross section which is about 2 fb at $\sqrt{S}=140$  GeV. The future EIC is expected to reach 100 fb$^{-1}$ integrated luminosity, hence we can expect at least 10 event signals. As we have discussed above, within a relevant kinematical range, this cross-section is resulted from a contribution of only CO state, $^{3}{S}{_1}^{[8]}$, so their measurement could provide a clean probe of CO mechanism, and hence the NRQCD framework in general. Using the cross-section data, one can fit the CO LDME $\langle 0\mid \mathcal{O}^{J/\psi}(\leftidx{^{3}}{S}{_1}^{[8]})\mid 0\rangle$, and hence it can also  provide a clean extraction of this LMDE \cite{Mehen:1996vx}.

\subsection{Upper Bound}\label{sec:upperbound}

In general, the polarized gluon TMDs are expected to satisfy the model independent positivity bounds, which are given in Ref.\cite{Mulders:2000sh}. By saturating the positivity bounds on the TMDs, we  estimate the allowed model independent upper bound of the asymmetries, which is defined as the maximum of the absolute value the asymmetry. From  Eqs.\eqref{eq:cos2phiT}-\eqref{eq:h1Tg-asy} we obtain the following upper bound relations on the absolute value of the $A^{\text{cos}2\phi_T}$ and  $A^{\text{cos}2(\phi_T-\phi_\perp)}$ asymmetries, 
\begin{align}\label{eq:ub}
\vert A^{\cos 2 \phi_\sT}\vert \leq 2\frac{ |{\cal B}_0|  }{{\cal A}_0}\,, \quad  \vert A^{\cos 2 (\phi_\sT-\phi_\perp)}\vert \leq 2\frac{ |{\cal B}_2|  }{{\cal A}_0}\,,
\end{align}
and the upper bound for the Sivers asymmetry, $A^{\sin(\phi_S-\phi_\sT)}$, becomes equal to one, while the upper bounds for $A^{\sin(\phi_S+\phi_\sT)}$ and $A^{\sin(\phi_S-3\phi_\sT)}$ are just half of the upper bound of $ A^{\cos 2 \phi_\sT}$. We show  the upper bounds  at two fixed values of $Q^2$, namely, $Q^2=10,20~\mathrm{GeV}^2$. We take the value of $K_\perp$ as the same order of $M_\psi$, i.e, $K_\perp=3~\text{GeV}\approx M_\psi$. We found that the above upper bounds of the asymmetries  are maximum at $y=0.1$ and $z=0.4$ for $\vert A^{\text{cos}2\phi_T}\vert$, while $\vert A^{\text{cos}2(\phi_T-\phi_\perp)}\vert$ is maximum at $z=0.7$. Moreover, the upper bounds of these asymmetries are independent of the center of mass energy, $\sqrt{S}$, as the coefficients ${\cal B}_i$ and ${\cal A}_i$ are independent of $\sqrt{S}$. The only dependence of $\sqrt{S}$ on the asymmetries comes through longitudinal momentum fraction, $x$, see Eq.~\eqref{dfun}. As we discussed above, in our case, the cross-section depends only on one LDME, that is  $\langle 0\mid \mathcal{O}^{J/\psi}(\leftidx{^{3}}{S}{_1}^{[8]})\mid 0\rangle$. This allows the asymmetries to be independent of the LDME as this cancels in the ratios. This makes the process considered here important for probing the gluon TMDs. \par
\begin{figure}[H]
	\begin{center} 
		\begin{subfigure}{0.5\textwidth}
		\includegraphics[height=7cm,width=8.5cm]{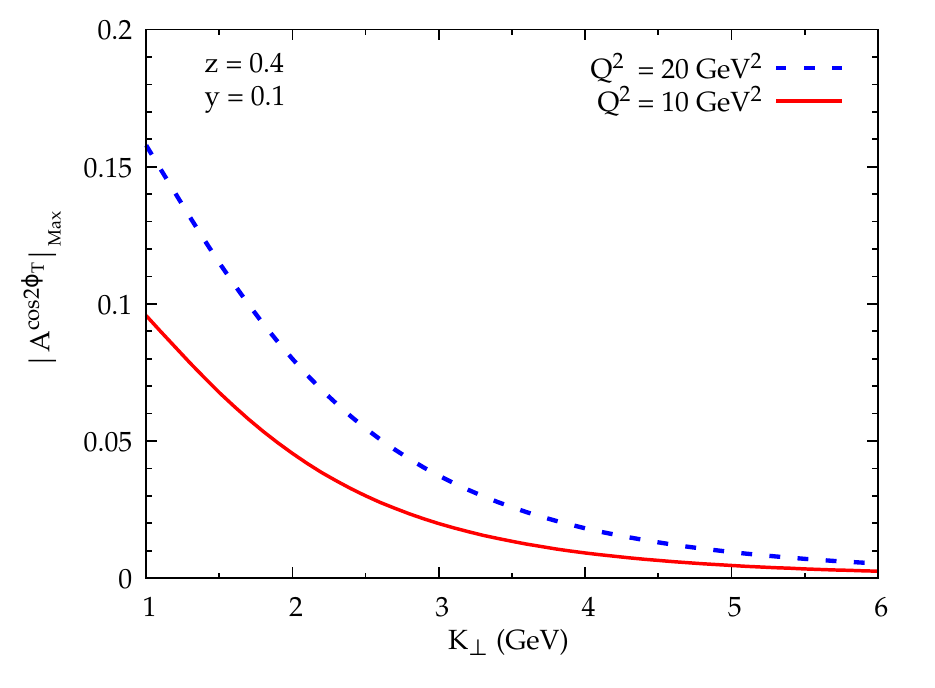}
			\caption{}
		\end{subfigure}
	    \begin{subfigure}{0.49\textwidth}
	    \includegraphics[height=7cm,width=8.5cm]{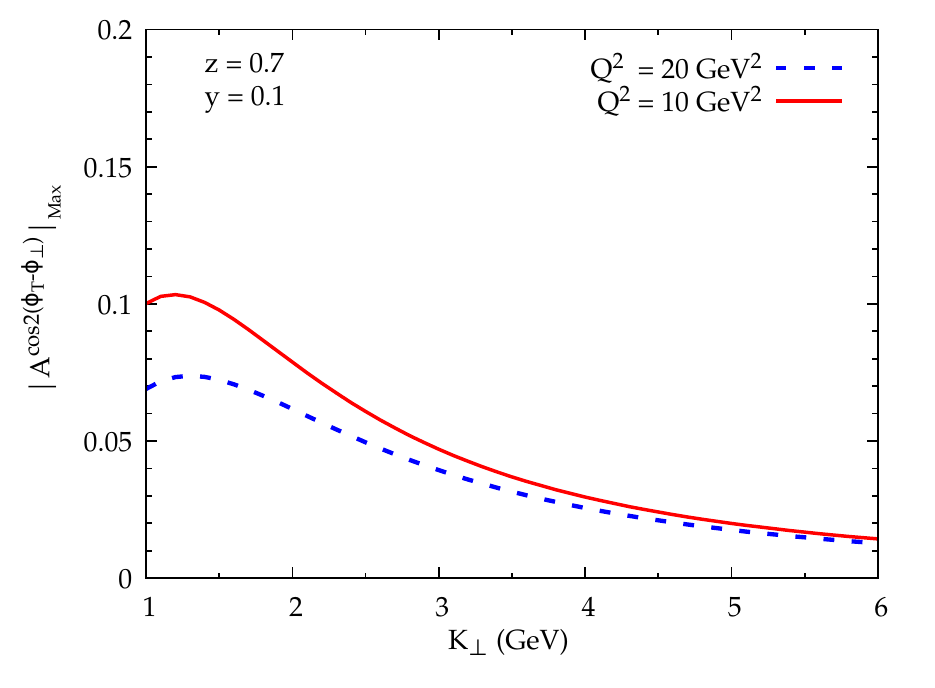}
	        \caption{}
	    \end{subfigure}
	\end{center}
	\caption{\label{fig:ub_vKp} Upper bound for the $A^{\text{cos}2\phi_T}$ (left panel) and $A^{\text{cos}2(\phi_{T}-\phi_\perp)}$ (right panel) azimuthal asymmetries in $e+p\rightarrow e +J/\psi+\gamma +X$ process as function of $K_\perp$ at fixed values of $y=0.1$, $z=0.4$ (left) and $z=0.7$ (right) for two values of $Q^2$.}
\end{figure}

In Fig.~\ref{fig:ub_vKp}, we show the upper bound of $A^{\text{cos}2\phi_T}$ (left panel) and $A^{\text{cos}2(\phi_T-\phi_\perp)}$ (right panel) as function of the transverse momentum of the outgoing particle $K_\perp$ for two values of the  virtuality of the photon  $Q^2=10,20~\mathrm{GeV}^2$. The dependence on $Q^2$ is  interesting, the magnitude of $A^{\text{cos}2\phi_T}$ increases with increasing $Q^2$, whereas $A^{\text{cos}2(\phi_T-\phi_\perp)}$ decreases with increasing $Q^2$. However, at  relatively large virtuality of  the photon, both these asymmetries do not show   significant dependency. Such behavior is more evident in the relatively low transverse momentum $K_\perp$ regions, where these asymmetries are quite sizable.\par
\begin{figure}[H]
	\begin{center} 
	    \begin{subfigure}{0.49\textwidth}
	    \includegraphics[height=7cm,width=8.5cm]{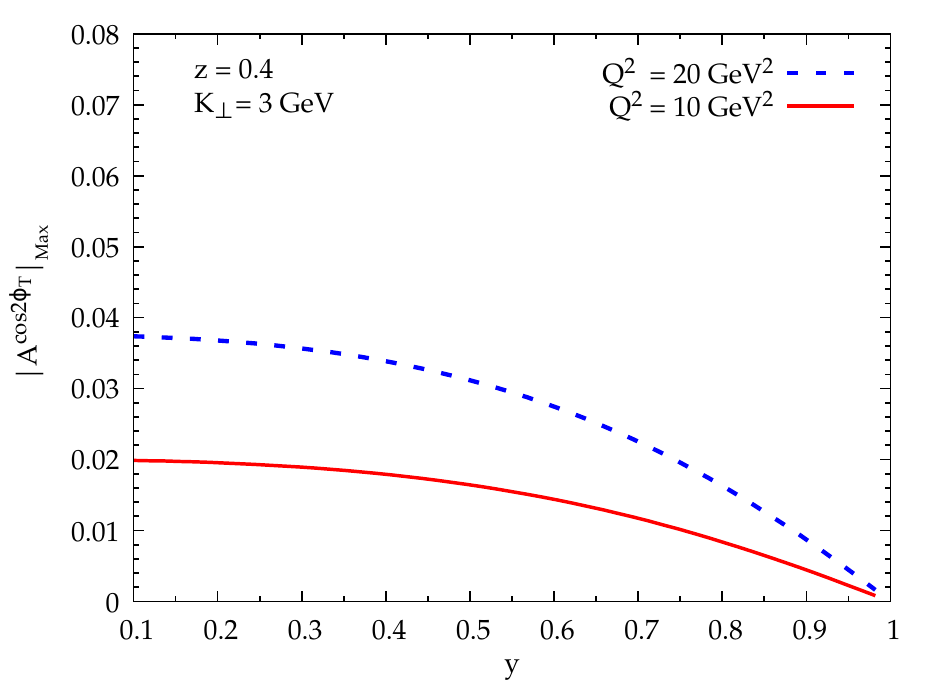}
	    \caption{}
	    \end{subfigure}
	    \begin{subfigure}{0.49\textwidth}
	    \includegraphics[height=7cm,width=8.5cm]{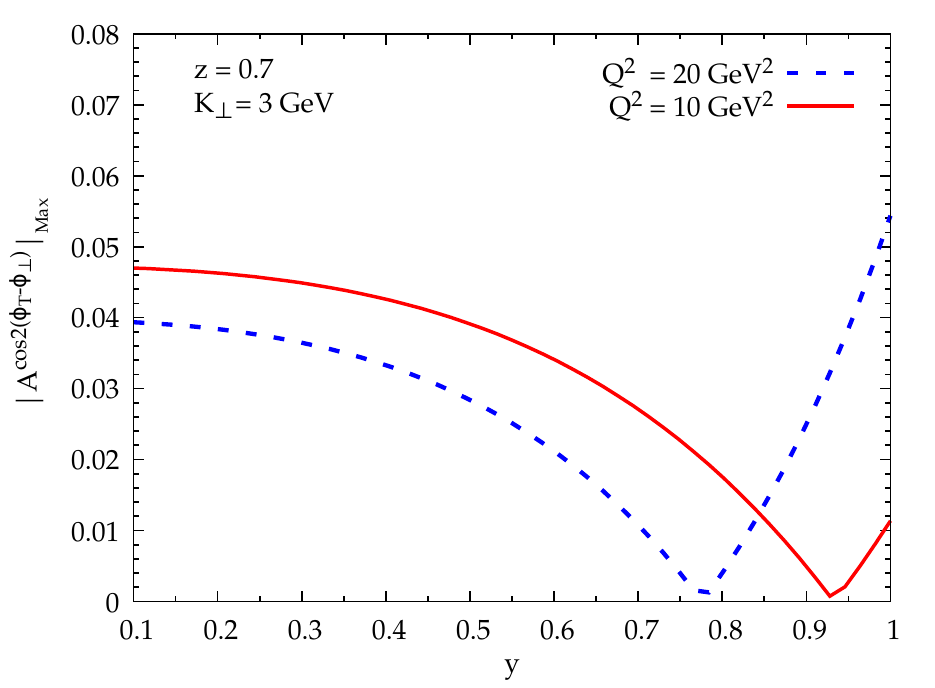}
        \caption{}	    
	    \end{subfigure}
	\end{center}
	\caption{\label{fig:ub_vy} Upper bound for the $A^{\text{cos}2\phi_T}$ (left panel) and $A^{\text{cos}2(\phi_{T}-\phi_\perp)}$ (right panel) azimuthal asymmetries in $e+p\rightarrow e +J/\psi+\gamma +X$ process as function of $y$ at fixed values of $K_\perp=3~\text{GeV}$, $z=0.4$ (left)  and $z=0.7$   (right) for two values of $Q^2$.}
\end{figure}

The $y$ dependence of the above upper bounds are shown in  Fig.~\ref{fig:ub_vy}. In the lower  $y$ ($y<0.3$) region, the nature of both these bounds are similar as both are approaching saturation in their magnitude, so we have considered $0.1<y$ as we do not see any significant change in their magnitudes beyond this. However, the nature of these two bounds are different, in the upper limit of $y\to 1$. The upper bound of $A^{\text{cos}2\phi_T}$ vanishes, which is expected as there is a factor of $(1-y)$ in the expression of  $\mathcal{B}_0$ coefficient, wherein only the longitudinal photon contributes, see the Eq.\eqref{eq:B0}. However, for $A^{\text{cos}2(\phi_T-\phi_\perp)}$, both transverse and longitudinal photons contribute, see $\mathcal{B}_2$ coefficient in Eq.\eqref{eq:B2}. As a result it vanishes before $y\to 1$, in particular, at $y\approx 0.8$ and $ 0.9$ for $Q^2=20$  and  $10$ GeV$^2$ respectively.\par
\begin{figure}[H]
	\begin{center} 
	    \begin{subfigure}{0.49\textwidth}
	    \includegraphics[height=7cm,width=8.5cm]{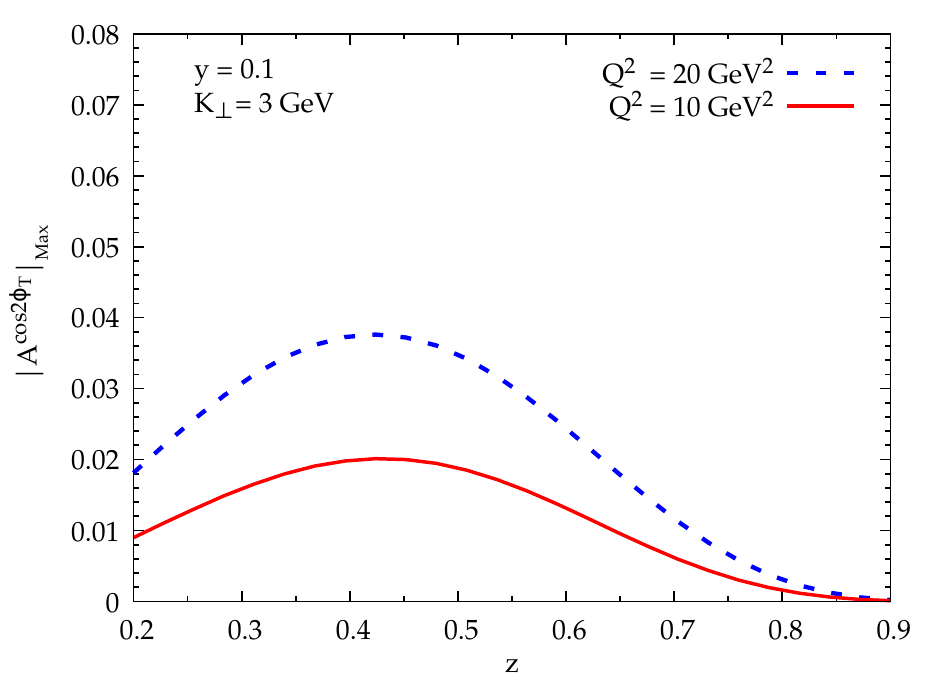}
	    \caption{}
	    \end{subfigure}
	    \begin{subfigure}{0.49\textwidth}
	    \includegraphics[height=7cm,width=8.5cm]{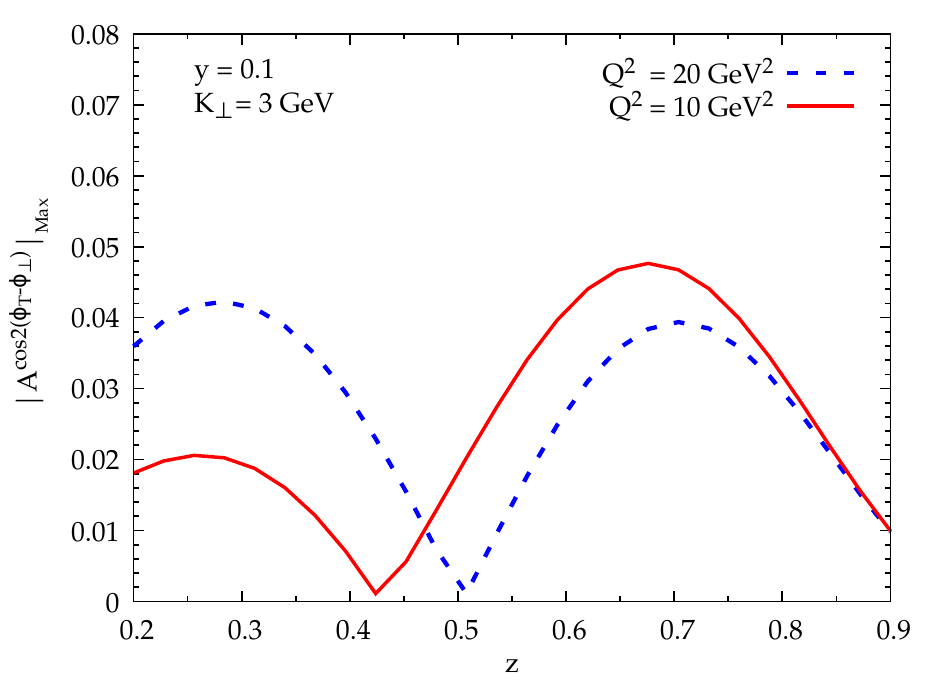}
        \caption{}	    
	    \end{subfigure}
	\end{center}
	\caption{\label{fig:ub_vz} Upper bound for the $A^{\text{cos}2\phi_T}$ (left panel) and $A^{\text{cos}2(\phi_{T}-\phi_\perp)}$ (right panel) azimuthal asymmetries in $e+p\rightarrow e +J/\psi+\gamma +X$ process as function of $z$ at fixed values of $y=0.1$ and $K_\perp=3~\text{GeV}$ for two values of $Q^2$.
	}
\end{figure}

In Fig.~\ref{fig:ub_vz}, we show the upper bounds of asymmetries as function of $z$. As discussed before,  we consider $z$ in  the region $0.3 < z < 0.9$. The asymmetry $A^{\text{cos}2\phi_T}$ vanishes as $z\to 0.9$. The reason being the coefficient $\mathcal{B}_0$ vanishes in that limit. However, the upper bound of $A^{\text{cos}2(\phi_T-\phi_\perp)}$  vanishes at $z\approx 0.4$ and 0.5 for $Q^2=10$ and $20~\text{GeV}^2$ respectively. The  nature of this asymmetry can be attributed to the contribution from both transversely and longitudinally polarized photons  in the modulation of the amplitude  $\mathcal{B}_2$.

\subsection{ Gaussian Parameterization of the TMDs}
One can parametrize  gluon TMDs to estimate predictions. These estimates are dependent on the parametrizations,  in contrast to the model-independent positivity bounds as discussed above. We adopt the following Gaussian parametrization for the linearly polarized gluon TMD $h_1^{\perp g}$ as given in Ref. \cite{boer2012linearly,boer2012polarized} 
\begin{eqnarray}
			\label{eq:gauss_h1p}
			h_1^{\perp g}(x,\bm{q}_\sT^2)=\frac{M_p^2f_1^g(x,\mu)}{\pi\langle q_\sT^2\rangle^2}\frac{2(1-r)}{r}e^{1-\frac{\bm{q}_\sT^2}{r\langle q_\sT^2\rangle}},
\end{eqnarray}
where, $M_p$ is the proton mass. $r(0<r<1)$ and the average intrinsic transverse momentum width of incoming gluon, $\langle q_\sT^2\rangle$, are parameters to this model. In our numerical estimation, we take $r=1/3$ and $\langle q_\sT^2\rangle=1~\text{GeV}^2$. For the unpolarized gluon TMD, $f_1^g$, we use the parametrization given in Eq.\eqref{eq:gauss_f1}.\par

\begin{figure}[H]
	\begin{center} 
		\begin{subfigure}{0.5\textwidth}
		\includegraphics[height=7cm,width=8.5cm]{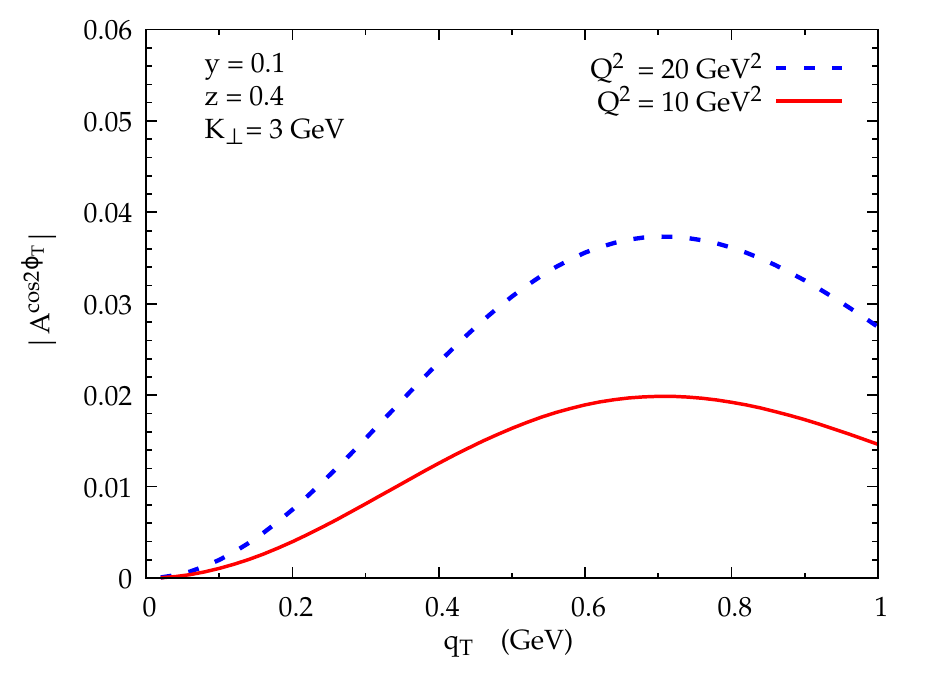}
			\caption{}
		\end{subfigure}
	    \begin{subfigure}{0.49\textwidth}
	    \includegraphics[height=7cm,width=8.5cm]{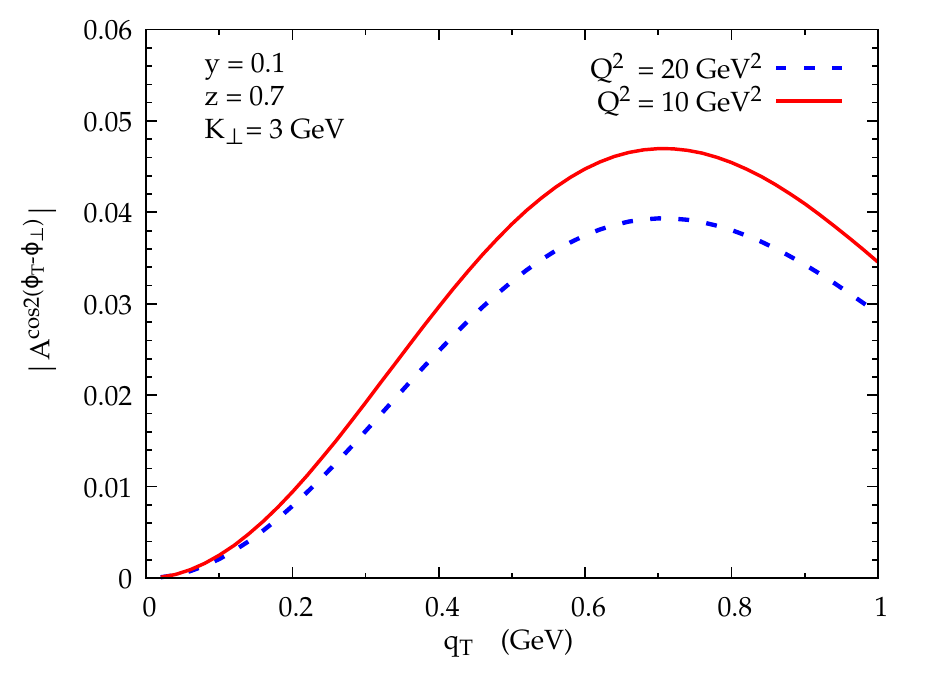}
	        \caption{}
	    \end{subfigure}
	\end{center}
	\caption{\label{fig:gauss_vqt_fix} Absolute values of $A^{\text{cos}2\phi_T}$ (left panel) and $A^{\text{cos}2(\phi_{T}-\phi_\perp)}$ (right panel) azimuthal asymmetries in $e+p\rightarrow e +J/\psi+\gamma +X$ process as function of $q_T$ at fixed values of  $K_\perp=3~\text{GeV},~y=0.1~\&~ z=0.4$ (left) and $K_\perp=3~\text{GeV},~y=0.1~\&~ z=0.7$ (right) for two values of $Q^2$. }
\end{figure}

In order to obtain model dependent $A^{\text{cos}2\phi_T}$ asymmetry, we have substituted Eqs.\eqref{eq:gauss_h1p}  and \eqref{eq:gauss_f1} for $h_1^{\perp g}$ and $f_1^g$ respectively in Eq.\eqref{eq:cos2phiT}. In the same way, $A^{\text{cos}2(\phi_T-\phi_\perp)}$ is obtained by using Eq.\eqref{eq:cos2phiT2phiP}.  In Fig.~\ref{fig:gauss_vqt_fix}, we show  $q_\sT$ dependent $A^{\text{cos}2\phi_T}$ (left panel) and $A^{\text{cos}2(\phi_T-\phi_\perp)}$ (right panel) azimuthal asymmetries, which are estimated for the same fixed values of  kinematic variables that were discussed in Sec.\ref{sec:upperbound}. By fixing the kinematics, the asymmetries become independent of $\sqrt{S}$, though they show a slight dependence on $\sqrt{S}$ via their dependence on $x$. Both these asymmetries have peak value at $q_\sT\approx 0.7$ which are about $4\%$ for $A^{\text{cos}2\phi_T}$ at $y=0.1,~z=0.4$ and $K_\perp=3$ GeV for $Q^2=20~\text{GeV}^2$, whereas $A^{\text{cos}2(\phi_T-\phi_\perp)}$ is estimated about $5\%$ at $y=0.1,~z=0.7$ and $K_\perp=3$ GeV for $Q^2=10~\text{GeV}^2$ .\par

\begin{figure}[H]
	\begin{center} 
	    \begin{subfigure}{0.49\textwidth}
	    \includegraphics[height=7cm,width=8.5cm]{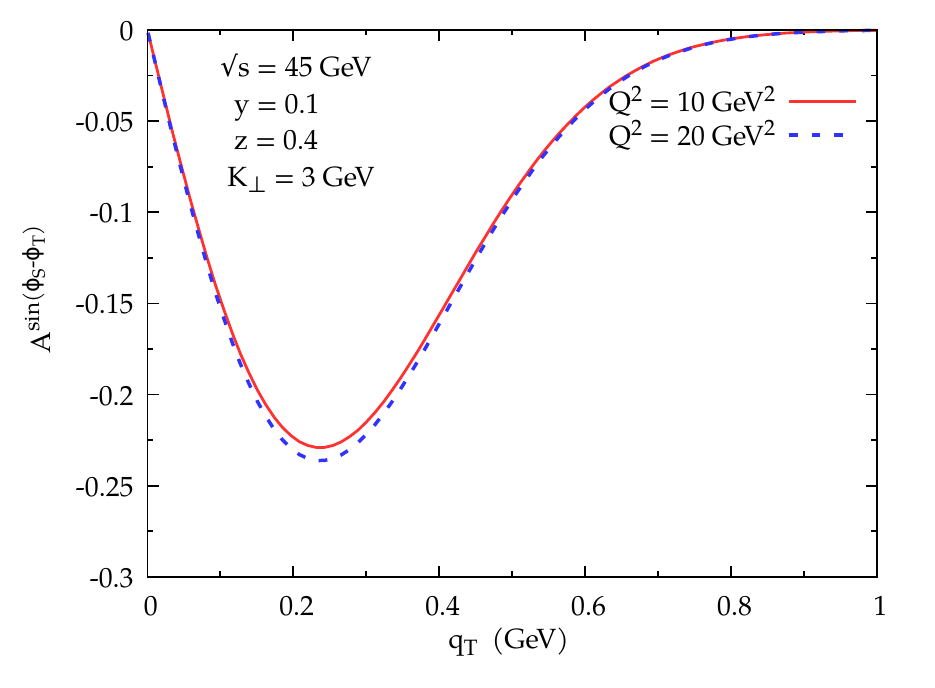}
	    \caption{}
	    \end{subfigure}
	    \begin{subfigure}{0.49\textwidth}
	    \includegraphics[height=7cm,width=8.5cm]{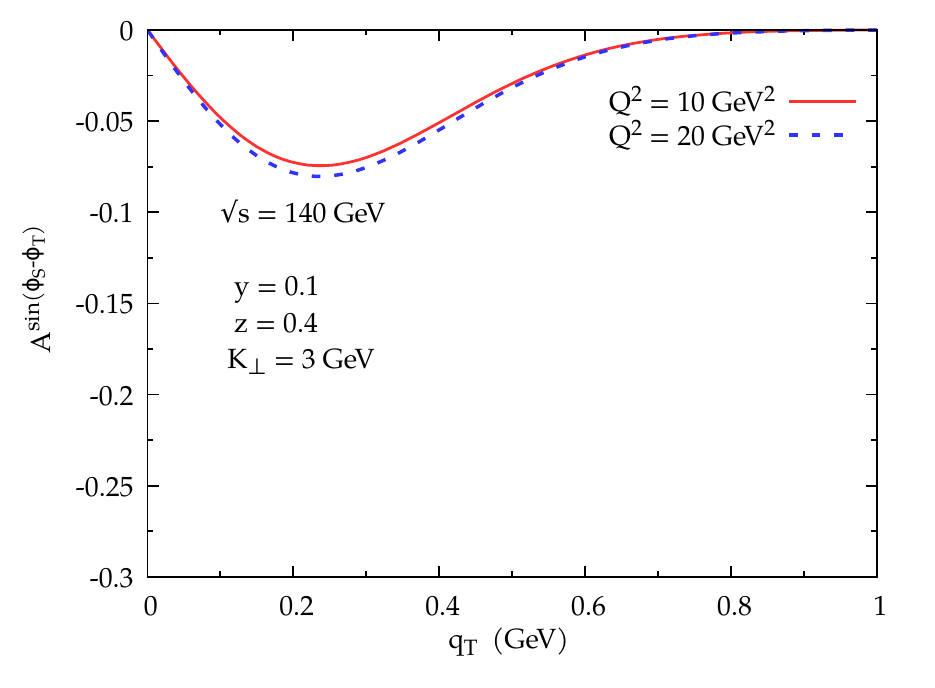}
        \caption{}	    
	    \end{subfigure}
	\end{center}
	\caption{\label{fig:gauss_sivers_vqt_fix} Sivers Asymmetry $A^{\text{sin}(\phi_{S}-\phi_T)}$ in $ep^{\uparrow}\rightarrow e+ J/\psi+\gamma +X$ process as function of $q_T$  at $\sqrt{s}=45~\text{GeV}$ (left) and $\sqrt{s}=140~\text{GeV}$ (right) for fixed values of $K_\perp=3$ GeV,  $y=0.1$ and $z=0.4$ and two different values of $Q^2$. }
\end{figure}

Similarly, we use the gluon Sivers function (GSF) $f_{1T}^{\perp g}$ parametrization  adopted in Ref.\cite{DAlesio:2018rnv} 
\begin{eqnarray}\label{eq:Sivers:par}
            \Delta^{N} f_{g / p^{\uparrow}}\left(x, q_\sT\right)= \left(-\frac{2|\bm{q}_\sT|}{M_P}\right)f_{1T}^{\perp g}\left(x, q_\sT\right)=2\frac{\sqrt{2 e}}{\pi} \mathcal{N}_{g}\left(x\right) f_{g / p}\left(x\right)  \sqrt{\frac{1-\rho}{\rho}} q_\sT \frac{e^{-\bm{q}_\sT^2 / \rho\left\langle q_\sT^2\right\rangle}}{\left\langle q_\sT^2\right\rangle^{3 / 2}}\,,
\end{eqnarray}

where
\begin{eqnarray}\label{eq:Sivers:Ng}
            \mathcal{N}_{g}\left(x\right)=N_{g} x^{\alpha}\left(1-x\right)^{\beta} \frac{(\alpha+\beta)^{(\alpha+\beta)}}{\alpha^{\alpha} \beta^{\beta}}\,,
\end{eqnarray}
and the extracted best fit parameters at $\left\langle q_\sT^2\right\rangle=1~\text{GeV}^2$ are 
\be
 N_g=0.25\,,\quad \alpha=0.6\,, \quad \beta=0.6\,,\quad \rho=0.1 \,.
\ee
To obtain model dependent Sivers asymmetry, we substituted    Eqs.\eqref{eq:Sivers:par} and \eqref{eq:gauss_f1} for $f_{1T}^{\perp g}$ and  $f_1^g$  respectively in Eq.\eqref{eq:f1t_sivers}. 
In Fig.~\ref{fig:gauss_sivers_vqt_fix}, the Sivers asymmetry $A^{\text{sin}(\phi_{S}-\phi_T)}$ is shown as a function of $q_\sT$ at $\sqrt{S}=45~\text{and}~140$ GeV. Within the above adopted Gaussian parametrization, it turns out that the Sivers asymmetry strongly depends on $\sqrt{S}$. This is due to the presence of a $x$ dependent term $\mathcal{N}_g(x)$ in the parametrization, see Eq.\eqref{eq:Sivers:par}.     
The estimated Sivers asymmetry is negative and is about $25\%$ and $7\%$ for $\sqrt{s}=45$ and 140 GeV respectively at fixed kinematic variables  $y=0.1$ and $z=0.4$ and $K_\perp=3$ GeV. The reason for large Sivers asymmetry at lower energy is that the magnitude of $\mathcal{N}_g(x)$, given in  Eq.\eqref{eq:Sivers:Ng}, is one order of magnitude larger than at the higher energy. Asymmetry hardly depends on the virtuality due to the presence of  the same coefficient $\mathcal{A}_0$ in the numerator and denominator of the Sivers asymmetry definition.

\section{Conclusion}\label{sec5}
In this work, we have presented a study of $J/\psi-\gamma$ pair production in  electron-proton collision within the kinematics of future proposed EIC. We have considered a kinematical scenario where the $J/\psi$ and photon($\gamma$) are almost back-to-back in the transverse plane. We have assumed the TMD factorization for this process and used the NRQCD framework for the $J/\psi$ production where only CO state, $^3S_1^{(8)}$, contributes. Thus, the only LDME present gets canceled in the asymmetry, which can be a good probe of the gluon TMDs.
Both the unpolarized and polarized cross-sections contains some specific azimuthal modulations. In particular, the $A^{\text{cos}2\phi_T}$ and $A^{\text{cos}2(\phi_{T}-\phi_\perp)}$ azimuthal asymmetries can probe the linearly polarized gluon TMD, whereas the $A^{\text{sin}(\phi_{S}-\phi_T)}$ can  probe the gluon Sivers TMD.  We have used the theoretical positivity bounds on the gluon TMDs that allow  us to estimate the model independent upper bound of these asymmetries. Apart from these model independent estimates, we have also estimated the transverse momentum dependent asymmetries using the Gaussian  parametrization of TMDs. Within the kinematics we have considered, the estimated model independent upper bounds on the asymmetries as well as the predictions made with using a model show some significant azimuthal asymmetries which would be very useful experimental probes at the EIC  to some of the still unknown gluon TMDs.

\section{Acknowledgement}
AM acknowledges the funding from Board of Research in Nuclear Sciences (BRNS), Govt.
of India, under sanction No. 57/14/04/2021-BRNS/57082. DC and RK acknowledge the funding from Science and Engineering Research Board under the Grant No. CRG/2019/000895.

\appendix
\section{Amplitude modulations }\label{appen}
We redefine the partonic Mandelstam variables as the following
\begin{eqnarray*}
 s=\hat{s}+Q^2,\quad u=\hat{u}+Q^2, \quad t=\hat{t}\,.
\end{eqnarray*}
The amplitude modulations are listed here
\begin{align}\label{A0:mod}
 \mathcal{A}_0&=-\frac{32(4\pi)^4 \alpha_s \alpha^3 e_c^4  Q^2  }{3 M_\psi s^2 y^2 (s+t)^2 (s+u)^2 (t+u)^2} \Bigg\{(1+(1-y)^2)\Big[2 Q^6 t^2 \left(s^2+t^2\right)-Q^4 t \big(s^3 (3 t-2 u)+3 s^2 t (t+u)\nonumber\\&+2 s t^2 (t-u)+2 t^3 (t+u)\big)+Q^2 \left(s^4 t^2+s u^2 (s-2 t) \left(s^2+t^2\right)-2
   s t u (s+t) \left(s^2+t^2\right)\right)\nonumber\\&-s^2 (s+t+u) \big(s^2 \left(t^2+t u+u^2\right)+s t u (t+u)+t^2 u^2\big)\Big]+ 16(1-y) Q^2  \Big[2 Q^4 t^2 \left(s^2-2 t^2\right)\nonumber\\&-2 Q^2 t \left(s^2-2 t^2\right) (s (t-u)+t (t+u))+s \left(u^2 \left(s^3+s t^2+4 t^3\right)+2 t^2 u (s+t) (s+2 t)+s t^2
   (s+t)^2\right)\Big]  \Bigg\}\,.
\end{align}
\begin{align}
 \mathcal{A}_1&=-\frac{64 (4\pi)^4 \alpha_s \alpha^3 e_c^4 M_\psi Q^2 t  }{3  s^2 y^2 (s+t)^2 (s+u)^2 (t+u)^2} (1-y)\left(Q^2 t+s u\right) \left(-2 Q^2 t^2+s^3+s^2 (t+u)\right)\,.
\end{align}
\begin{align}
 \mathcal{A}_2&=-\frac{64(4\pi)^4 \alpha_s \alpha^3 e_c^4 Q^3 \sqrt{t \left(Q^2 t+s u\right)}  }{3 M_\psi s^2 y^2 (s+t)^2 (s+u)^2 (t+u)^2} \sqrt{1-y}\, (2-y) \Big[-4 Q^4 t^3+Q^2 t \big(s^3+s^2 (t+u)+2 s t (2 t-u)\nonumber\\
 &+4 t^2 (t+u)\big)+s u \left(s^2+2
   t^2\right) (s+t+u)\Big]\,.
\end{align}
\begin{align}\label{eq:B0}
 \mathcal{B}_0&=-\frac{32 (4\pi)^4 \alpha_s \alpha^3 e_c^4 M_\psi Q^4  t^2 }{3  s^2 y^2 (s+t)^2 (s+u)^2 (t+u)^2}(1-y)  \left(Q^2 t^2-s^3-s^2
   (t+u)\right)\,.
\end{align}
\begin{align}
 \mathcal{B}_1&=\frac{32(4\pi)^4 \alpha_s \alpha^3 e_c^4 M_\psi Q^3 t  \sqrt{t \left(Q^2 t+s u\right)}}{3   s^2 y^2 (s+t)^2 (s+u)^2 (t+u)^2}\sqrt{1-y}\, (2-y)  \left(-2 Q^2 t^2+s^3+s^2 (t+u)\right)\,.
\end{align}
\begin{align}\label{eq:B2}
 \mathcal{B}_2&=\frac{16(4\pi)^4 \alpha_s \alpha^3 e_c^4 M_\psi Q^2  t  }{3 
    s^2 y^2 (s+t)^2 (s+u)^2 (t+u)^2} \Big[(1+(1-y)^2)\left(Q^2 t+s u\right) \left(2 Q^2 \left(s^2+t^2\right)-s^2 (s+t+u)\right)\nonumber\\
    &- 4(1-y)  Q^2 \left(s^2-2 t^2\right)   \left(Q^2 t+s u\right)\Big]\,.
\end{align}

\begin{align}
 \mathcal{B}_3&=-\frac{64  (4\pi)^4 \alpha_s \alpha^3 e_c^4 M_\psi Q^3 t  \sqrt{t \left(Q^2 t+s u\right)}}{3   s^2 y^2 (s+t)^2 (s+u)^2 (t+u)^2}\sqrt{1-y}\, (2-y)  t  \left(Q^2 t+s u\right)\,.
\end{align}

\begin{align}
 \mathcal{B}_4&=-\frac{32 (4\pi)^4 \alpha_s \alpha^3 e_c^4 M_\psi Q^2  t^2 }{3  s^2 y^2 (s+t)^2 (s+u)^2 (t+u)^2} (1-y)\left(Q^2 t+s u\right)^2\,.
\end{align}

\bibliographystyle{apsrev}
\bibliography{references}
\end{document}